\documentclass[twocolumn]{article}

\usepackage{arxiv}
\usepackage{geometry}
\usepackage{hyperref}
\usepackage{cite}
\usepackage{enumitem}
\setlist[itemize]{noitemsep, nolistsep}
\usepackage{graphicx}
\graphicspath{{figures/}{pictures/}{images/}{./}} 

\usepackage{amsmath}
\usepackage[ruled,vlined]{algorithm2e}
\usepackage{algpseudocode} 
\usepackage{array}
\usepackage{microtype}
\usepackage{caption}
\usepackage{booktabs}                  
\usepackage[usenames,dvipsnames,table]{xcolor}
\newcommand{\annot}[1]{
\iftrue 
\textbf{[#1]}
\fi
}

 \newcommand{\comment}[1]{}
 
\title{Nanouniverse:\\ Virtual Instancing of Structural Detail and Adaptive Shell Mapping}
\usepackage{authblk}
\author[ ]{Ruwayda~Alharbi}
\author[ ]{  Ond\v{r}ej~Strnad}
\author[ ]{Markus~Hadwiger}
\author[ ]{Ivan~Viola}

\affil[ ]{King Abdullah University of Science and Technology (KAUST), Saudi Arabia.\newline E-mails: \{ruwayda.alharbi $\vert$ ondrej.strnad $\vert$ markus.hadwiger $\vert$ ivan.viola\}@kaust.edu.sa.}

\date{2022}                   
\begin{document}
\twocolumn[
  \begin{@twocolumnfalse}
    \maketitle
    \begin{abstract}
Rendering huge biological scenes with atomistic detail presents a significant challenge in molecular visualization due to the memory limitations inherent in traditional rendering approaches. In this paper, we propose a novel method for the interactive rendering of massive molecular scenes based on hardware-accelerated ray tracing. Our approach circumvents GPU memory constraints by introducing virtual instantiation of full-detail scene elements. Using instancing significantly reduces memory consumption while preserving the full atomistic detail of scenes comprising trillions of atoms, with interactive rendering performance and completely free user exploration.
We utilize coarse meshes as proxy geometries to approximate the overall shape of biological compartments, and access all atomistic detail dynamically during ray tracing. We do this via a novel adaptive technique utilizing a volumetric shell layer of prisms extruded around proxy geometry triangles, and a virtual volume grid for the interior of each compartment. Our algorithm scales to enormous molecular scenes with minimal memory consumption and the potential to accommodate even larger scenes. Our method also supports advanced effects such as clipping planes and animations. We demonstrate the efficiency and scalability of our approach by rendering tens of instances of Red Blood Cell and SARS-CoV-2 models theoretically containing more than 20 trillion atoms.

    \end{abstract}
    
  \end{@twocolumnfalse}
]


\section{Introduction}\label{sec:introduction}

\begin{figure*}
     \centering
    \includegraphics[width=\linewidth, alt={A view of a city with buildings peeking out of the clouds.}]{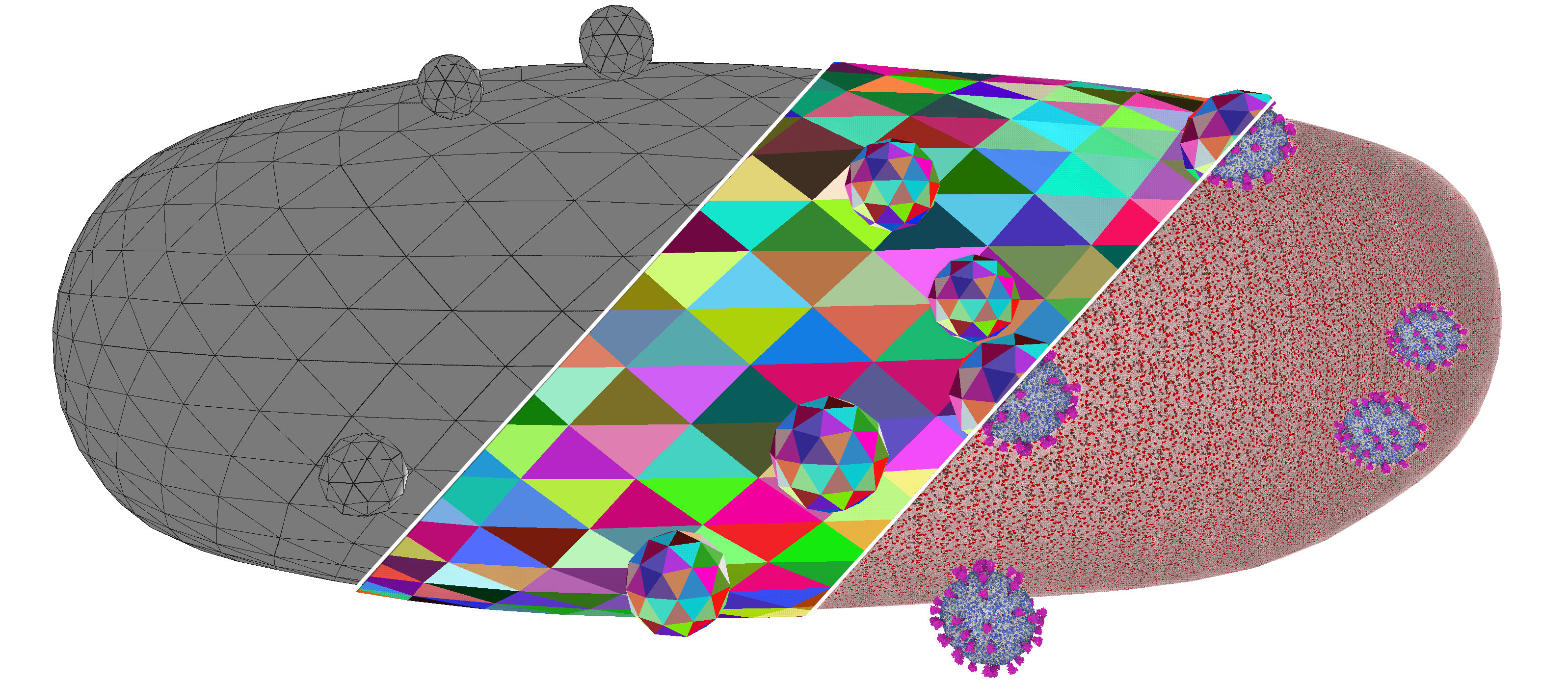}
  \caption{Red Blood Cell (RBC) occluded by SARS-CoV-2 particles, with full atomistic detail. (Left) The proxy geometry comprising the overall scene structure; (Middle) The intermediate acceleration structures for ray tracing; (Right) The resulting model with on-the-fly instancing, rendered at interactive rates via ray tracing at the level of individual atoms.}
  \label{fig:teaser}
 \end{figure*}

 In molecular visualization, representing atomistic models of microscale biological systems poses a major challenge due to the sheer number of atoms they contain. Already a tiny SARS-CoV-2 particle consists of two dozen million atoms. In the most common representation describing molecular structures, each atom is represented by a sphere. Depending on the chemical element, each sphere is characterized by a specific radius. A naive representation of such a scene would require at least four IEEE float values (for position and radius) for each atom, requiring almost 400~MB of memory to describe the structure of a single viral particle. Going further, even just a single Red Blood Cell (RBC) contains more than 1.2 trillion atoms, which is five orders of magnitude more than in a single viral particle. Such structures therefore would require dozens of terabytes or more to represent a single RBC. 
 
 However, biological systems are highly repetitive in nature. They most frequently contain four chemical elements: Carbon, Nitrogen, Oxygen. and Hydrogen. These elements are present in water, in aminoacids which form proteins, in nucleobases which form genetic molecules, and in lipids that form the membranes of these biological entities. The composition of higher-level structural elements, like proteins, genetic molecules, polysaccharides, or membranes, is also highly repetitive. For example, there are only 20 aminoacid types that form the entire variety of proteins, and there are only five nucleobases that form genetic molecules. Therefore, even if biological systems contain an enormous number of atoms, the repetitive nature of their buildup offers an opportunity to represent biological structures effectively, by smartly \emph{instancing} identical structural elements in a scene many times.
 
 Instancing is a technique in computer graphics for rendering multiple copies of the same geometry with minimal effort. The instancing process involves two distinct types of objects: The first is the \emph{geometric object}, which defines the geometry to be instanced. This object is defined in model space without any specific reference to the environment in which it will be rendered. The second type is the \emph{instance}, which represents a ``copy'' of the geometric object but only stores attributes such as the model-to-world transformation matrix. This scheme allows visualization techniques to render many instances of a geometric model while storing its underlying geometric data only once in GPU memory. The attributes of geometry instances must either be stored in a buffer or computed dynamically during rendering. Several prior works in molecular visualization exploit this idea of instancing to reduce the footprint of atomistic models for biological systems~\cite{Lindow-2012-Interactive-Rendering-Of-Materials,LeMuzic-2015-cellVIEW,Nanomatrix}. A single RBC contains around 518 million instances of molecular structures. Simply storing the position, rotation, and type of each instance requires 29 bytes of memory. This means that under the most ideal circumstances, the entire model requires approximately 15 GB of GPU memory. However, in practice additional memory overhead is incurred by the rendering algorithm, e.g., BVH data for ray tracing can require 3.5 times the raw data size~\cite{RTXpkd}. No prior technique is capable of rendering a single RBC with full detail on a GPU with 24 GB of memory.
 
It is an elegant property of ray tracing that it enables instancing in a natural and very efficient manner. If we have a ray that we want to intersect with a transformed instance, we can instead intersect an inverse-transformed ray with the untransformed geometric object~\cite{Marschner-and-Shirley-2016-Fundamentals-Of-Computer-Graphics-Book}.

In this paper, we develop this idea of instancing in ray tracing into a novel multi-level instancing approach for massive atomistic models of biological systems. At the highest level, our scenes comprise textured triangle meshes, which act as coarse proxy geometries for ray tracing and represent various biological entities such as viruses or cells. Starting from the proxy geometries, we capture the repetitive pattern of each type of biological entity by a \emph{mesostructure}, which is a set of Wang tiles for the membrane (the outer part), and a set of Wang cubes for solubles (the inner parts). The \emph{nanostructures} are then the individual molecules that form the mesostructures. During ray tracing, we virtually instantiate mesostructures and nanostructures in the scene by computing instance transformation matrices on the fly, and then transforming rays from the global scene to the local coordinate system of each instance using the inverse of each instance's transformation matrix. This avoids having to explicitly place complex atomistic geometry into the scene. Moreover, only the scene elements that are actually visible are checked for intersection. With this latter property we further drastically reduce memory requirements for representing biological systems, while maintaining interactive framerates without any level-of-detail scheme that might compromise rendering accuracy.

The main contributions of this paper are:
\begin{itemize}[noitemsep]
\item Multi-level virtual instancing to minimize memory footprint;
\item A novel three-level nested acceleration structure for ray tracing;
\item An adaptive volumetric \emph{shell space} to render mesostructures protruding from the base mesh of a proxy geometry;
\item A \emph{core space} to encompass the proxy geometry's interior;
\item An evaluation of visual, computational, and memory allocation aspects of our technique using scenes with trillions of atoms.
\end{itemize}

\section{Related Work}
\begin{figure*}[t!]
    \centering
    \includegraphics[width=0.95\linewidth]{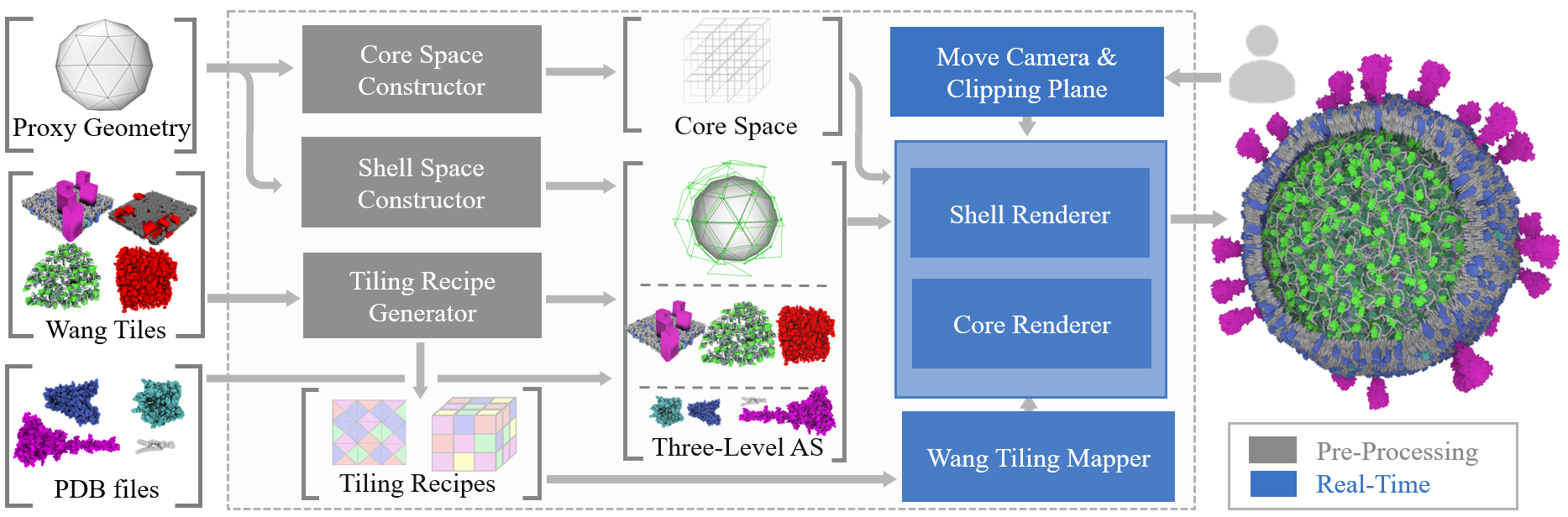}
    \vspace{0.5mm}
    \caption{
    \textbf{The Nanouniverse system} constructs the scene out of \emph{proxy geometries} (top left), which are then filled with Wang tiles (middle left), comprising Wang squares for their volumetric shells (\emph{shell space}) corresponding to membranes, as well as Wang cubes for their 3D interior (\emph{core space}). Wang tiles are populated with the atomistic detail of instances of proteins from PDB files (bottom left). Wang tiling recipes are created in a pre-processing step. In addition, a three-level acceleration structure (AS) is built to accelerate rendering using ray tracing. During rendering, the interiors of all Wang tiles are instantiated virtually and rendered with full atomistic detail.}
   \label{fig:Nanouniverse}
   \vspace{-4mm}
\end{figure*}

\paragraph*{\textbf{Large-scale Molecular Visualization.}}
Various tools for molecular visualization, such as VMD~\cite{VMD}, VIAMD~\cite{VIAMD} or PyMOL~\cite{PyMOL}, are available. However, these tools are designed to handle molecules with up to thousands of atoms and struggle when dealing with datasets exceeding tens of millions of atoms~\cite{Knoll-2013-Ray-Tracing-and-Volume-Rendering-Large-Molecular-Data-on-Multi-Core-and-Many-Core-Architectures}. 
Megamol~\cite{Grottel-2015-MegaMol, OSMegamol} is a visualization framework tailored to interactive visualization of large particle-based datasets, capable of rendering up to 100 million atoms at interactive framerates, equivalent to the scale of a virus or a small bacterium in biology. Lindow et al.~\cite{Lindow-2012-Interactive-Rendering-Of-Materials}  have first presented interactive visualization of large-scale biological data, comprising several billion atoms. They utilized a grid structure for each protein, storing all atoms on the GPU as a 3D voxel, and employed instancing to replicate proteins in the scene. 
Falk et al.~\cite{Falk-2013-Atomistic-Visualization-of-Mesoscopic} built upon this foundation by enhancing depth culling and refining the rendering method, incorporating an implicit Level of Detail (LOD) approach. Their optimizations enabled them to render 25 billion atoms at 3.6 FPS. In contrast, Le Muzic et al.~\cite{LeMuzic-2014-Illustrative-Visualization-of-Molecular-Reactions-using-Omniscient-Intelligence,LeMuzic-2015-cellVIEW} introduced a novel approach utilizing a simplified LOD scheme that circumvents the need for grid-based supporting structures. Instead, they employed the tessellation shader to dynamically inject sphere primitives into the rasterization pipeline for each molecular instance. This set a benchmark with 250 copies of an HIV virus model in blood plasma at 60 FPS. This scene contains 16 billion atoms, with each replica comprising approximately 60 million atoms. Our proposed approach goes far beyond by visualizing biological models with trillions of atoms. It is important to highlight that all the techniques mentioned earlier rely on procedural impostors to represent atoms, effectively simplifying the geometry and accelerating rendering~\cite{Tarini06,Michel20}. However, no prior method is able to visualize a single RBC. Nanomatrix~\cite{Nanomatrix} uses view-guided construction, partitioning the scene into boxes dynamically filled with geometric representations close to the camera, and image textures farther away. While it successfully renders huge models, it only represents a part with atomistic detail. Another drawback lies in the abrupt transition between image and geometry representations, leading to noticeable discrepancies in intricate details such as protrusions at contours. Our approach addresses this issue and renders the entire model at full detail, while also integrating advanced techniques such as cutaway views. Virtual instantiation liberates us from GPU memory constraints to render models of any size, provided that their proxy geometries and Wang tiles fit in memory. 
\paragraph*{\textbf{Mesostructure Procedural Modeling.}} 
Modeling at the mesoscale presents significant challenges due to the dense and diverse nature of molecular structures in terms of size and shape. Various techniques have been developed to address this, such as cellPACK~\cite{cellPACK}, which utilizes packing algorithms for constructing mesoscale models. However, the computational demands of packing molecules make this process time-consuming, with model assembly ranging from minutes to hours depending on complexity. 
To enable the interactive creation of complex 3D content, research efforts have shifted towards the exploration of parallelization techniques~\cite{greuter2003real}. This strategy focuses on efficiently distributing computation and memory tasks across graphics hardware, thereby improving overall performance. Klein et al.\cite{Klein-2019-Parallel-Generation-and-Visualization-of-Bacterial-Genome-Structures,Tobias-2018-Instant-Construction} introduce the concept of ``instant construction,'' which aims to swiftly generate mesoscale models. They utilize a series of GPU-based population algorithms to produce a variety of biological structures. Inspired by the texture synthesis literature, they leverage Wang tiling~\cite{wang1961proving} to accelerate the construction of mesostructures on the surface of quad-based meshes. However, their instant construction is limited by the available GPU memory. 
Nanomatrix~\cite{Nanomatrix} expands upon this work by extending it to triangle meshes and a view-guided approach. They utilize Wang square tiles and box tiles
to fill the scene with biological structure. Despite the scalability of Nanomatrix's construction algorithm, its renderer was restricted by hardware limitations. 
Therefore, Nanomatrix renders only the close portion of the scene in full detail while utilizing a textured image to represent the rest of the scene. In this paper, we propose a method that can render the entire scene in full detail. Unlike Nanomatrix and other on-the-fly construction algorithms~\cite{Steinberger2014,greuter2003real}, our approach does not allocate any buffers for constructed instances but computes and renders them on the fly, avoiding dynamic memory management during rendering.

\paragraph*{\textbf{Displacement Mapping and Shell Mapping.}}
Various tessellation-free techniques have been developed for real-time rendering of surface details. Patterson et al.~\cite{Patterson-Inverse-Displacement-Mapping} proposed inverse displacement mapping, which incorporates surface details during color texturing. This became popular, and additional displacement mapping methods were proposed, such as parallax mapping and relief mapping. For more on these methods, Szirmay-Kalos and Umenhoffer's survey~\cite{Displacement-Mapping-on-the-GPU-State-of-the-Art} offers a comprehensive overview.
Thonat et al.~\cite{Thonat-2021-Tessellation-Free-Displacement-Mapping-for-Ray-Tracing,Thonat-2023-RMIP} introduced an interactive displacement mapping approach for hardware ray tracing. 
However, displacement mapping struggles to accurately represent overhangs due to its reliance solely on height information from the texture map, lacking additional data about the geometry of such structures. Porumbescu et al.~\cite{Porumbescu2005Shellmaps} introduced shell mapping to improve the realism of surface representations in objects, similar to displacement mapping. However, shell mapping surpasses displacement mapping in terms of its expressive capabilities, enabling the generation of intricate mesostructures, including overhangs, in a layer between the base mesh and its offset mesh, which is termed a shell. The algorithm maps a 3D volume onto the mesh surface by subdividing each prism forming the shell into three tetrahedra. During rendering, when the rays intersect the tetrahedron, entry and exit points define a ray segment that is then converted from world space to texture space using the barycentric coordinates of the tetrahedra. Subsequently, the ray progresses within the shell's texture space to find the intersection. The piecewise linear approximation of prism decomposition into tetrahedra may result in significant aliasing artifacts. Jeschke et al.~\cite{Jeschke-2007-Interactive-Smooth-and-Curved-Shell-Mapping} tackled this problem by introducing a smooth curved shell mapping, which employs ray marching in texture space by solving a cubic equation. Our approach uses the texture space to define transformation matrices for mesostructures, while ray tracing of mesostructures operates in object space. We use shell space to instantiate protein models on the surface meshes of proxy geometries, with a new definition of shell space to ensure an accurate representation of biological data. Furthermore, we extend this idea to encompass the proxy geometry's interior as core space, facilitating the rendering of mesostructures in the interior of the proxy geometry. While the shell space is defined as the union of triangular prisms, we define the core space as the union of rectangular prisms (i.e., boxes).
\paragraph*{\textbf{RT Acceleration Structures.}}
The latest NVIDIA GPUs accelerate ray tracing through RT cores~\cite{NVIDIA-2018-Turing-GPU-Architecture-whitepaper},
supported by APIs such as OptiX~\cite{OptiX}, Vulkan~\cite{VULKAN}, and Microsoft DXR~\cite{MICROSOFT-DXR}. Acceleration structures (AS) are crucial for ray tracing performance, with GPUs supporting bounding volume hierarchies (BVH). GPUs, however, allow accessing only two levels of RT AS: the bottom-level structure (BLAS), containing the geometric primitives, and the top-level structure (TLAS), composed of instances associated with transformation matrices and references to one of the BLASes~\cite{RayTracingGems}. NVIDIA OptiX 7 introduced support for multi-level instancing, enabling the creation of hierarchical instancing structures (TLAS). This feature allows instances themselves to contain further groups of instances (TLASes), forming a multi-level instancing hierarchy~\cite{OWL-OptiX7}. This enables replicating a group of instances several times within a scene, but the transformation matrices of each instance at each level must be defined and stored prior to rendering. Our approach utilizes a multi-level geometric hierarchy, where BLASes point to additional groups of BLASes. Instances are virtually instantiated from these geometries by computing transformation matrices on the fly. 
RT cores incorporate a specialized unit dedicated to bounding box intersection testing, essential for BVH traversal~\cite{NVIDIA-2018-Turing-GPU-Architecture-whitepaper}. During ray traversal, when the ray intersects BLAS geometry, programmable shaders are invoked, determining ray behavior at the hit point. However, the current hardware implementation does not provide control over ray traversal at the TLAS instance level or the BVH's internal node level. Incorporating a BLAS as an intermediate node in our multi-level hierarchy provides us with more control over the ray traversal. For example, if an instance is clipped away by a clipping plane, our multi-level AS allows terminating ray traversal at the instance level, eliminating the need to trace the ray against the geometry of that instance.


\section{Technical Overview}


\begin{figure*}[t!]
    \centering
    \includegraphics[width=\linewidth]{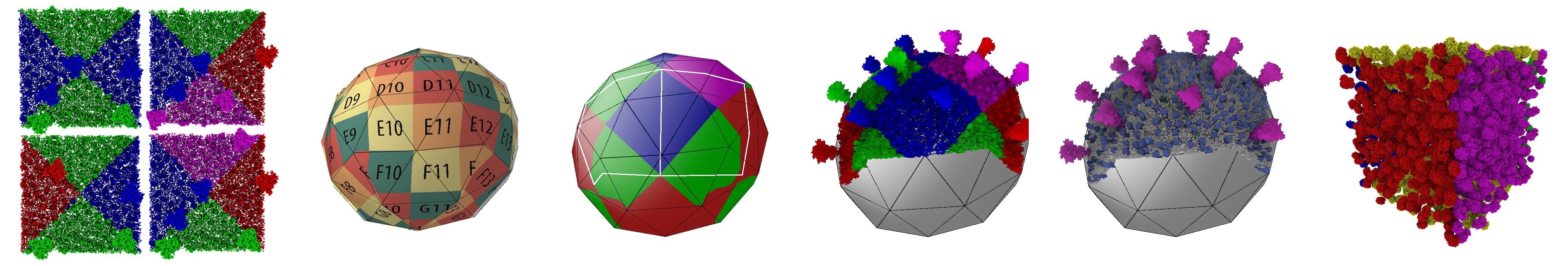}
    \caption{\textbf{Geometric Wang tiles} are either (left-most) Wang square tiles, or (right-most) Wang cube tiles. Both contain full atomistic detail. Middle part of the figure illustrates the process of mapping Wang square tiles to the corresponding uv coordinates in the Wang tiles texture.  }
   \label{fig:WangTiles}
   \vspace{-4mm}
\end{figure*}

The aim of Nanouniverse is to render complex biological scenes containing unprecedented amounts of atoms. To achieve this goal, due to their large size our models cannot be stored explicitly in memory.
We therefore decompose the scene into three major basic building blocks (see~\autoref{fig:Nanouniverse}): (1) Proxy geometries, (2) Wang tiles, and (3) Proteins.

Corresponding to these three main building blocks, our system uses three main input elements:
The first input are the \emph{proxy geometries} that define the geometry of biological compartments. The scene may contain several instances of various biological structures, and for each structure type there is an associated proxy geometry given as a triangle mesh with a low triangle count.
The second input type are geometric \emph{Wang tiles}, which comprise a set of collision-free molecule instances, where each has an associated transformation matrix. Wang tiles can be generated from larger 3D patches as described by Klein et al.~\cite{Tobias-2018-Instant-Construction}. These biological models are constructed based on biological {\it rules}, for example using Mesocraft~\cite{Mesocraft}, which defines the concentration of various molecules in a biological structure, and the principles that spatially characterize their relations. Our approach uses two types of geometric Wang tiles: \emph{Wang square tiles}, to represent membranes, and \emph{Wang cube tiles}, to represent soluble components. Both are illustrated in \autoref{fig:WangTiles}.
In the following, we will use the term  \emph{Wang tile} as a general term encompassing both types. Each Wang tile is defined as a square (or cube) with edges (or faces) encoded using a (conceptual) color, which restricts how the tiles can be placed during the tiling process to form a seamless tiling. In the pre-processing step, a description of such an arrangement is created, which we call a \emph{tiling recipe}. A tiling recipe is defined as a 2D/3D table of pointers that refer to one of the existing Wang tiles.
The third and final major type of input is the structure of all types of molecules that will comprise the scene, given as \emph{PDB files}. These files are available at the Protein Data Bank  (\href{https://www.wwpdb.org/}{wwpdb.org}).

\begin{figure}[b!]
    \centering
    \includegraphics[width=\linewidth]{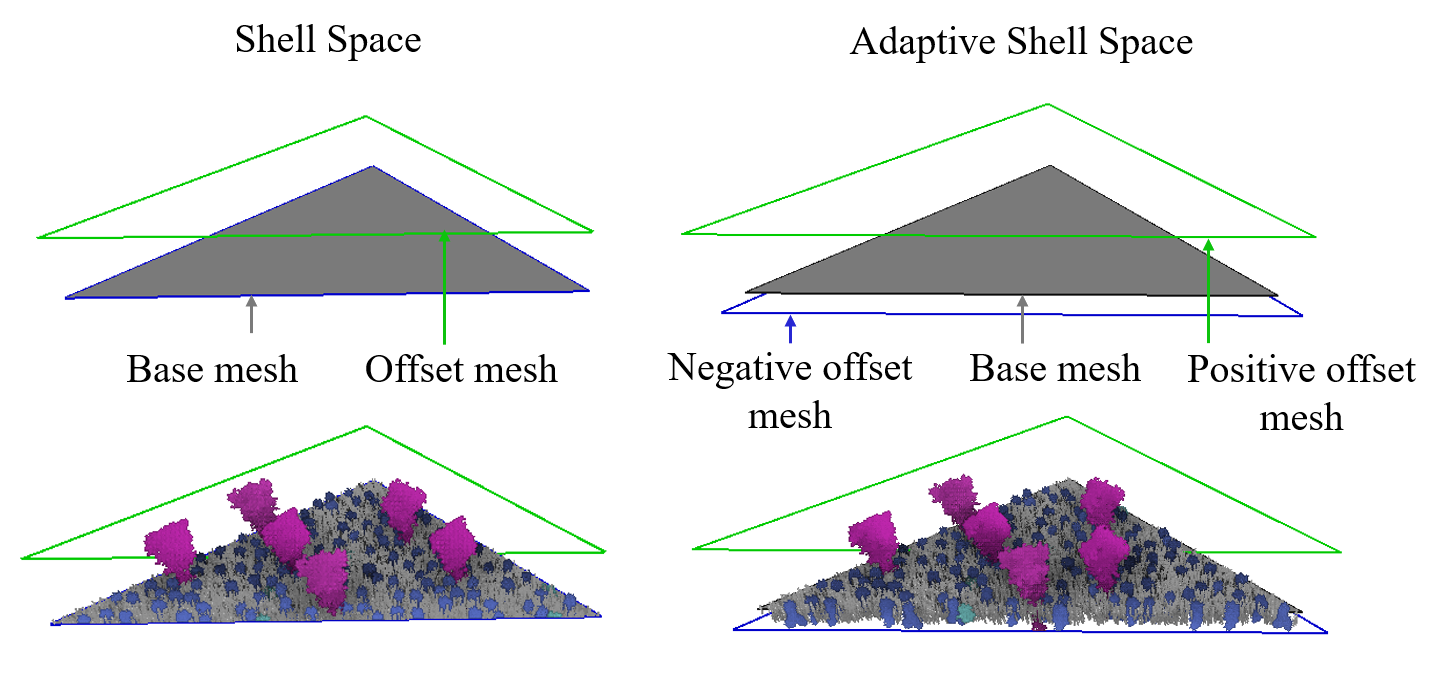}
    \caption{\textbf{The shell space} (left) is the volumetric layer between a base mesh and an offset mesh. The \textbf{adaptive shell space} (right) is the layer between positive and negative offset meshes. The bottom row shows both types of shell spaces filled with mesostructures with atomistic detail.}
   \label{fig:shellspace}
\end{figure}

The biological structures can be classified based on their scale as \emph{microstructures},  \emph{mesostructures}, and \emph{nanostructures}.
The rendered model is constructed on the fly, with per pixel ray-mesh intersections. To accelerate virtual construction and rendering, we separate the data into multiple acceleration structures. 
Two types of mesostructures are virtually placed within the scene with respect to the proxy geometry: (1) The membrane, which is a thin barrier of a lipid bilayer and proteins that separates the interior of a biological entity from its external environment. (2) The soluble components which consist of a set of molecules that fill the interior of biological compartments. Each of these structures uses different construction and rendering techniques.

\begin{figure*}[t!]
    \centering
    \includegraphics[width=\linewidth]{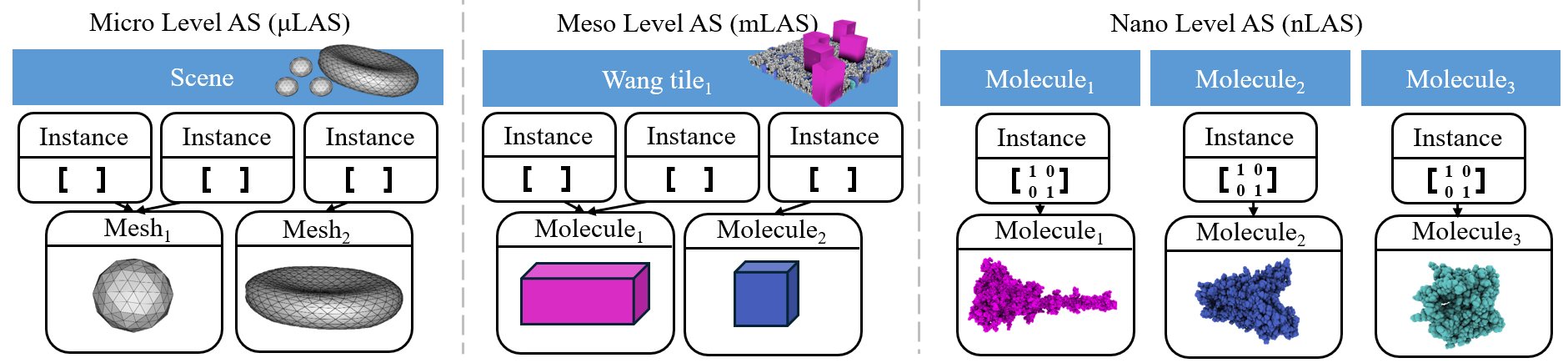}
    \caption{\textbf{Three levels of acceleration structures (AS) used in Nanouniverse.} Micro Level AS ({$\mu$LAS}) store the main geometry in the scene. Meso Level AS ({$m$LAS}) contains Wang tiles and cubes. Nano Level AS ({$n$LAS}) are designed for protein instances data. }
   \label{fig:three-level-AS}
   \vspace{-4mm}
\end{figure*}

\begin{figure}[b!]
    \centering
    \includegraphics[width=0.8\linewidth]{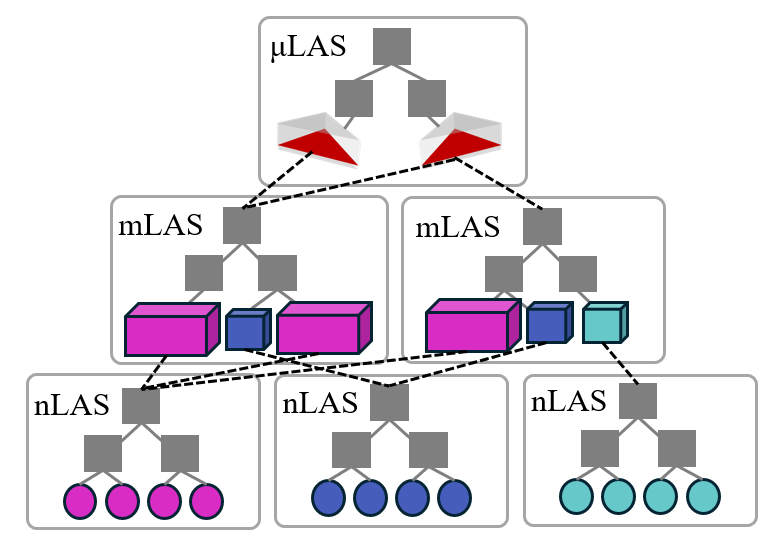}
    \caption{\textbf{The three-level acceleration structure (AS) tree.} The Micro Level AS ({$\mu$LAS}) represents the highest level in the tree, followed by the Meso Level AS ({$m$LAS}). Then, Nano Level AS ({$n$LAS}) which represents the lowest level in the tree. A single geometry representing a molecule can be instantiated as a part of a Wang tile, and the Wang tile can be instantiated to form the scene.}
   \label{fig:multi-level-instancing}
\end{figure}

\paragraph*{\textbf{Shell Space and Adaptive Shell Space.}}
Because our scene consists of proxy geometries given as low-poly meshes, and we want to be able to render details possibly protruding from the mesh also around contours of the mesh, we draw complex mesostructures that represent membranes in a volumetric layer between a \emph{base mesh} and its \emph{offset mesh}. We call this layer the \emph{shell space} or, simply, the shell (see \autoref{fig:shellspace}).
The shell is constructed in the pre-processing phase by extruding the base mesh towards an offset mesh along its normals at each vertex. The shell is defined as the union of prisms generated by the triangles of the base mesh along with their corresponding offset mesh~\cite{Porumbescu2005Shellmaps,Kautz2007InteractiveSmoothandCurvedShellMapping}. As our target is to visualize molecular biological structures, we extend the basic definition of the shell space and propose an \emph{adaptive shell space} that extrudes each triangle individually on height and width based on the mesostructures that are mapped to that triangle.

\paragraph*{\textbf{Core Space.}}
We further extend the concept of the shell space with a \emph{core space} to be able to represent the proxy geometry's interior. This facilitates visualization of mesostructures located further within the biological compartment than the base or offset mesh. While the shell space is defined as the union of triangular prisms, the core space is defined as the union of rectangular prisms, or, simply, \emph{boxes}. We define the core space by uniformly subdividing the proxy geometry’s axis-aligned bounding box (AABB). This constructs a set of non-overlapping 3D axis-aligned boxes that together form the \emph{core grid}, representing the core space. Due to the grid's regularity, there is no need to construct an explicit data structure for the core grid's boxes. All essential positional information can directly be derived from the grid AABB, box size, and grid dimensions. We refer to these properties as the \emph{core grid metadata}.

\paragraph*{\textbf{Acceleration Structures.}}
In order to facilitate interactive visualization at high framerates, we aim to use the GPU ray tracing acceleration structures of NVIDIA RTX GPUs to represent the scene.
The RTX acceleration structure is a bounding volume hierarchy (BVH) that is GPU hardware-accelerated. It is completely abstracted by the NVIDIA drivers and it is not possible to access any of its information outside the drivers. During ray traversal, NVIDIA allows the developer to access only the leaf nodes of that hierarchy to make decisions such as accepting/ignoring an intersection. Because we need better control of the ray traversal process, we therefore propose partitioning the scene into several RTX acceleration structures (AS) (see \autoref{fig:three-level-AS}):
The \emph{Micro Level AS ($\mu$LAS)}, which defines the scene and contains all the polygonal mesh instances that define the shape, size, and position of the biological entity.
The \emph{Meso Level AS ($m$LAS)}\footnote{Within this article the prefix $m$- does not stand for milli but for meso unit.}, an AS for every Wang tile, which contains the description of the position, rotation, and type of each molecular instance in Wang tile model, and finally the \emph{Nano Level AS ($n$LAS)}, an AS for each molecular type, which contains the position of a molecule's atoms (represented of spheres with variable radii based on the type of the atom). 

These three levels of RTX acceleration structures together form a \emph{three-level tree} system (see~\autoref{fig:multi-level-instancing}), where the leaf node of each level points to the root of one or more RTX acceleration structures in the next lower level. In this tree system, the micro level AS ($\mu$LAS) represents the root. Every leaf node in $\mu$LAS, which represent an extruded prism, is pointing to the root of one or more Wang tiles in $m$LAS. Every leaf node on $m$LAS, which represents the molecule's axis-aligned bounding box (AABB), which is a rectangular box that encompasses the entire molecule's atoms, points to the root node of that molecule in $n$LAS. By using this three-level tree system, a single geometry representing a molecule can be instantiated multiple times as a part of a Wang tile, and the Wang tile can be instantiated multiple times to form the scene.

We need to define how to map from the leaf node of one level to the root of the next lower level. Using the molecule Id, we can directly map from $m$LAS to  $n$LAS (i.e., from the molecule's AABB to its AS). However, to map from $\mu$LAS to $m$LAS (i.e., from prism to Wang tiles), we need to know which Wang tiles from the tiling recipe should be placed in that prism. For this, we use a \emph{Wang tile mapper}, which is a mapping function introduced in previous work~\cite{Nanomatrix}. For mapping the Wang squares to the triangle mesh of the proxy geometry, the idea is to create the tiling recipe in such a way that it should be big enough to cover the entire $(u,v)$~texture coordinates. Then, based on the triangles' texture coordinates, we can project any triangle to the tiling recipe and identify the list of Wang tiles that should be placed in that triangle. We refer to this list as the \emph{replication area}. For mapping Wang cubes, the tiling recipe is expected to be big enough to cover the entire core space, such that there is mapping for every box to a corresponding Wang cube. We refer the reader to the original work for more details~\cite{Nanomatrix}.

\paragraph*{\textbf{Ray Traversal of Three-Level Trees.}}
During the rendering stage, Nanouniverse renders the scene by performing a ray traversal on its {\it three-level tree}. As the ray moves through the scene, the algorithm first traverses the structure represented by the $\mu$LAS. After the intersection of the ray and a prism is found, it idles the traversal on $\mu$LAS and switches the traversal to the $m$LAS. The traversal traces against a Wang tile at which the prism intersection is pointing to. To minimize the memory footprint, the Nanouniverse algorithm instantiates the instances virtually and computes the transformation matrices on the fly rather than storing them. Therefore, before issuing the traversal, we compute the transformation matrix that describes the transformation matrix of the Wang tile to the respective triangle of the mesh in the scene and use it to inverse the ray.
Once the ray hits a molecule on $m$LAS, we idle the traversal again and switch to $n$LAS. Before starting the traversal, we compute the transformation matrix that describes how that molecule should be positioned in the scene and use this transformation matrix to inverse the ray. Then, 
we traverse the ray in the lowest level of the hierarchy which is $n$LAS. Once there is a ray hit event detected, the respective leaf node becomes a candidate for ray intersection. At the end, the closest hit is recorded.  Inversing the ray allows Nanouniverse to store the geometric description of mesostructures/nanostructures only once in the GPU memory and virtually instantiates them when required during ray traversal.

\paragraph*{\textbf{Per-Structure Renderers.}}
Our models consist of two different types of mesostructures. Therefore, our approach incorporates two dedicated renderers--one for each type of structure. The \emph{shell renderer} is specifically designed for rendering membrane proteins by placing Wang square tiles in the mesh's shell space (the corresponding prisms), while the \emph{core renderer} manages the rendering of soluble components by placing Wang cubes in the mesh's core space (the core grid). Both perform the three-level ray traversal with minor differences. The shell renderer performs a ray-prism intersection test in $\mu$LAS while the core renderer performs a ray-AABB intersection test in $\mu$LAS. At the end of every frame draw call, the resulting image from both renderers is composited based on the depth. Both renderers operate at interactive rates, allowing the user to freely navigate the scene by changing the camera position or visibility settings.

\section{Pre-processing Phase}
We construct the prisms for the shell space and the grid metadata for the core space in a pre-processing step. The resulting data structures are uploaded to the GPU and remain unchanged during the interactive rendering stage. 

\subsection{Adaptive Shell Space Construction}
On the highest level, our scene consists of proxy geometry given as low-poly meshes. To achieve our goal of rendering intricate detail, including molecules protruding from the proxy geometry, we use a shell mapping-based approach. We render the biological membrane by drawing complex mesostructures with atomistic detail in the volumetric layer constituting the shell space (see \autoref{fig:shellspace}). As our aim is to visualize molecular biological structures, we refine the concept of the shell space and introduce the adaptive shell space. The adaptive shell is an automatically generated volumetric layer corresponding to each triangle of the proxy geometry that facilitates the visualization of molecular biological structures inside the shell layer.

As biological molecules can be located beneath or above the base mesh of the proxy geometry, we establish two distinct offset meshes. The first offset mesh arises from the extrusion of the base mesh along its vertex normals, which is referred to as the \emph{positive-offset mesh}. The second offset mesh is constructed by extruding the base mesh in the opposite direction of its vertex normals, and is referred to as the \emph{negative-offset mesh}. We define the adaptive shell space as the layer between these two offset meshes.
Tracing the ray in the shell layer is expensive. To optimize shell space traversal, we assign distinct offsets for each triangle, calculated based on the height along the $ \pm y$-axis of the largest molecule in the mesostructures that is mapped to that triangle based on its texture coordinates. Moreover, as biological structures are densely packed with molecules, we calculate the side offset based on the width of the largest molecule that might be placed on its edge. Subsequently, overlapping prisms are constructed by extruding the vertices of the offset meshes in the direction of the center-to-vertex vector. Mesostructures and nanostructures are placed inside this prism with respect to the texture coordinates of the respective triangles.

\subsection{Core Space Construction}
To render the soluble, our approach involves drawing the mesostructures within the core space of the proxy geometry. We apply uniform spatial subdivision to the proxy geometry's axis-aligned bounding box (AABB) to create a 3D grid referred to as the \textit{core grid}. The AABB is partitioned into non-overlapping axis-aligned \textit{boxes} with identical extents. Since a single Wang cube will be placed in each box, the grid is partitioned in such a way that the box size equals the size of a Wang cube. The resulting core grid is characterized by its AABB, the number of boxes along each axis, which defines the grid dimensions, and by the size of each individual box. Due to the uniformity of the grid, there is no need to construct an explicit data structure for the core grid's boxes. All geometric positions can be derived directly from the grid metadata. For example, for any 3D point $p_{xyz}$, the coordinates of a box $b_{ijk}$ within the grid can be calculated by dividing the world coordinates by the size of the box.
\begin{equation}
\label{eq:cijk_eq}
 b_{ijk} = (p_{xyz} - grid.min_{xyz})/b.size_{xyz} 
\end{equation}
We can also compute the minimum position of a box $b.min_{xyz}$ by multiplying the box's coordinates $b_{ijk}$ by box size and then add into the result the minimum position of the grid.
\begin{equation}
\label{eq:c_min_eq}
 b.min_{xyz} = (b_{ijk}\times b.size_{xyz}) + grid.min_{xyz} 
\end{equation}

\subsection{Three-level Acceleration Structure Tree}
\label{sec:three-level-AS}
Acceleration structures (AS) are the core ingredient of ray tracing with high performance. NVIDIA implemented this component in hardware and exposes two-level AS to developers: the BLAS contains the geometric primitives, whereas the TLAS contains one or more instances of BLASes with a transformation matrix~\cite{RayTracingGems}.

Inspired by this design, we propose three-level acceleration structures that provide us with greater control during ray traversal (refer to \autoref{fig:three-level-AS}). The first level is a singleton $\mu$LAS, which represents the scene. Within this AS, the BLAS describes mesh vertices and indices, where the prism represents the lowest primitive. The TLAS describes mesh instances, each accompanied by a world-space transformation matrix and a pointer to the corresponding mesh in the BLAS. The second level is the $m$LAS, which represents the Wang tiles and Wang cubes. Each Wang tile has its own instance of $m$LAS. Within this AS, the BLAS describes the molecule's AABB, which is the lowest primitive. The TLAS in this AS describes the molecular instances forming the Wang tile, with each instance accompanied by a object-space transformation matrix and a pointer to the corresponding molecule in the BLAS. Storing the Wang tile instances with object-space transformation matrices is important for reusing these tiles and virtually instantiating them in the scene. The third level is the $n$LAS, which contains all types of molecules to be rendered in the scene. Each molecule has its own instance of  $n$LAS. Within this AS, the BLAS describes the object-space positions of molecule's atoms, where the atom acts as the lowest primitive. In the TLAS, we place a single instance with an identity matrix as the transformation matrix. We choose to place the atoms representation in the BLAS because the atoms overlap significantly and it is the best practice for NVIDIA RTX Ray Tracing~\cite{nvidia_RTX_best_practices}.

The three levels of acceleration structures constitute a hierarchical tree system. Each leaf node at $\mu$LAS, representing an extruded prism, is linked to the root of one or more Wang tiles at $m$LAS. Similarly, each leaf node at  $m$LAS, representing molecule axis-aligned bounding boxes (AABBs), is linked to the root nodes of that molecule at  $n$LAS. These inter-level connections are computed during the ray traversal process. Establishing a method for transitioning from a leaf node at one level to the root of the next lower level is important. The molecule Id facilitates a direct linkage from  $m$LAS to  $n$LAS, connecting the molecule's AABB to that molecule's Acceleration Structure (AS). However, transitioning from $\mu$LAS to  $m$LAS, from prism to Wang tiles, is done based on the \textit{Wang tile mapper} described below. Based on the tiling recipe and prism's base triangle texture coordinates, the \textit{Wang tile mapper} determines which Wang tiles should be placed on that prism.

\begin{table}[t]
\centering
\caption{Symbols used in this paper}
\begin{tabular}{p{0.20\linewidth}p{0.70\linewidth}}
\hline
Symbol & Explanation\\
\hline
{\textbf{$n_{uv}, n_{xyz},$ $ n_{\mu\gamma}, n_{\mu\gamma\lambda}$}}& the position of molecular instance $n$ in texture/world or tile's object-space coordinates.\\ 
\textbf{$n_{\theta}$}& the rotation of an instance inside tile.\\
\textbf{$g_{uv},g_{xyz}$}& the center of a Wang tile in texture/world coordinates.\\
\textbf{$b_{ijk}$}&  column $(i)$, row $(j)$, and layer $(k)$ which represent the box $b$ location within the core grid.\\

\hline
\end{tabular}
\label{table:symbols}
\end{table}

\subsection{Wang Tile Mapper}
\label{sec:TilingRecipe_WangTileMapper}
For the placement and visualization of molecular structures at the correct spatial locations, we need a method that maps Wang squares and cubes to the prisms of shell space, or the boxes of core space, respectively. Previous work, Nanomatrix~\cite{Nanomatrix}, introduced a construction algorithm that transforms molecular instances from Wang square coordinates to the texture coordinates of a mesh, and from Wang cube coordinates to the AABB of a mesh. Nanouniverse uses the same Wang tile mapper to link $\mu$LAS and $m$LAS in its three-level AS tree. In this section, we briefly explain the Wang tile mapper algorithm, as this is a crucial part of our rendering algorithm. 

The method requires a triangular mesh as the input, where each vertex is associated with texture coordinates that fall within the range of $[0,1]$. In the preprocessing step, several components are prepared. 
As we aim to instantiate mesostructures inside the prisms during rendering, a mapping of a set of Wang squares to the proxy geometry's base mesh is needed. Therefore, in the pre-processing phase, we run the Wang tilling algorithm to generate the tiling recipe, which is a 2D array that contains indices of Wang squares that represent a valid Wang tiling arrangement. The tiling recipe is designed to be large enough to cover the full texture coordinate space (i.e., from $[0, 1]$) so that for any triangle in $uv$-space we are able to determine which Wang squares are projected onto that particular triangle. The same algorithm is extended to 3D to handle Wang cubes. 
The Wang cube tiling recipe is a 3D array that is large enough to cover the full core grid coordinates so that for any box a corresponding Wang cube can be determined. To parallelize the mapping process, Nanomatrix's Wang tile mapper uses a sliding window approach. It defines the maximum number of tiles within the tiling recipe that is selected for processing for any triangle in the mesh, this 2D region or window is called the \emph{replication area}. The mapper uses the largest triangle of the mesh to compute the size of the replication area during pre-processing. Later, during rendering, the $min_{uv}$ coordinates of the triangle define the origin of the replication area within the tiling recipe. From the replication area, we determine how many tiles are projected from the tiling recipe to the one particular triangle. For more details, we refer the reader to the original paper~\cite{Nanomatrix}.

\section{Nanouniverse Renderer}
\begin{figure*}[h]
    \centering
    \includegraphics[width=0.95\linewidth]{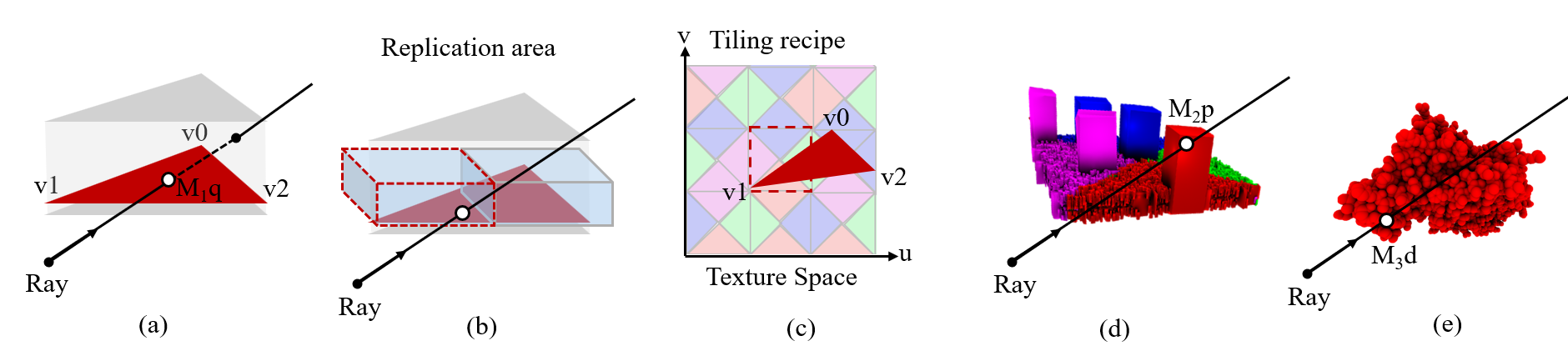}
    \caption{Illustration of Shell Space Renderer algorithm. (a) Illustration of a ray intersecting a prism. The AABB of tiles area traced (b) and a respective Wang square tile is determined (c). Then the ray continues to AABB of $n$LAS (d) and resulting atom of the instance of rendered (e).  }
   \label{fig:mambrane_renderer}
\end{figure*}

Our goal is to visualize large scenes with trillions of atoms at interactive rates. Moreover, we aim to keep the memory footprint as low as possible so the final scene can in the future reach over hundred of trillions of atoms, which is the average size of a human cell. Therefore, our scene uses geometric mesh as proxy geometries to represent the general shape of the biological compartment without any details. The details are added during the rendering process. 

Two types of mesostructures are virtually placed into the scene with the respect to each proxy geometry mesh: the membrane components, which shape the outer structure of a compartment and are oriented perpendicular to it, and the soluble components which fill the compartment. Our approach incorporates a dedicated renderer for each type: A {\it shell space renderer} for the membrane, and a {\it core space renderer} for solubles. Both renderers perform ray traversal on the three-level AS tree that is illustrated in~\autoref{fig:three-level-AS}. 
The main difference between the shell renderer and the core renderer is that the former performs ray-prism intersection tests at $\mu$LAS, while the latter performs ray-AABB intersection tests.

\subsection{Shell Space Renderer}
\autoref{fig:mambrane_renderer} explains the shell rendering algorithm with a simple example. We cast a ray $a+tb$ into the scene that has a single prism. The ray traversal starts in $\mu$LAS, looking for a ray-prism intersection. If there is a hit at point $q$  (\autoref{fig:mambrane_renderer}  (a)), the traversal continues to $m$LAS. Otherwise, the resulting color of the pixel is background. For the hit $q$, we identify first which AS in $m$LAS the ray has to further traverse. Furthermore, we compute a transformation matrix $M_1$ that is important for the correct virtual placement of the mesostructure to the scene. 

To identify the respective mesostructure at $m$LAS, we employ the texture coordinates of the triangles of the base proxy mesh. 
We use the vertices $(v_0, v_1, v_2)$ of the base triangle defining the intersected prism as the input into \emph{Tile Mapper} and returns the \emph{replication area}; a 2D list of Wang tiles that are projected from the tiling recipe onto that triangle.
From the pre-processing phase, the dimension of the \emph{replication area} is known. It is ($1\times2$) in our example (\autoref{fig:mambrane_renderer}  (b)). The \emph{Tile Mapper} uses the $uv$-coordinates of vertices to define $uv$ position $g_{uv}$ for every tile in the replication area. The algorithm tests the ray intersection with all of them and record the closest atom hit (if any), which could be done sequentially or in parallel, as we explain in Section~\ref{sec:opt-replication-area}. For every tile in the replication area, we obtain the 3D world position of the center of the tile $g_{xyz}$ by interpolation. This results in the translation matrix of the tile. Furthermore, we compute the rotation matrix from the triangle normal, tangent, and bi-tangent vectors as well. 

\begin{figure}[t]
    \centering
    \includegraphics[width=\linewidth]{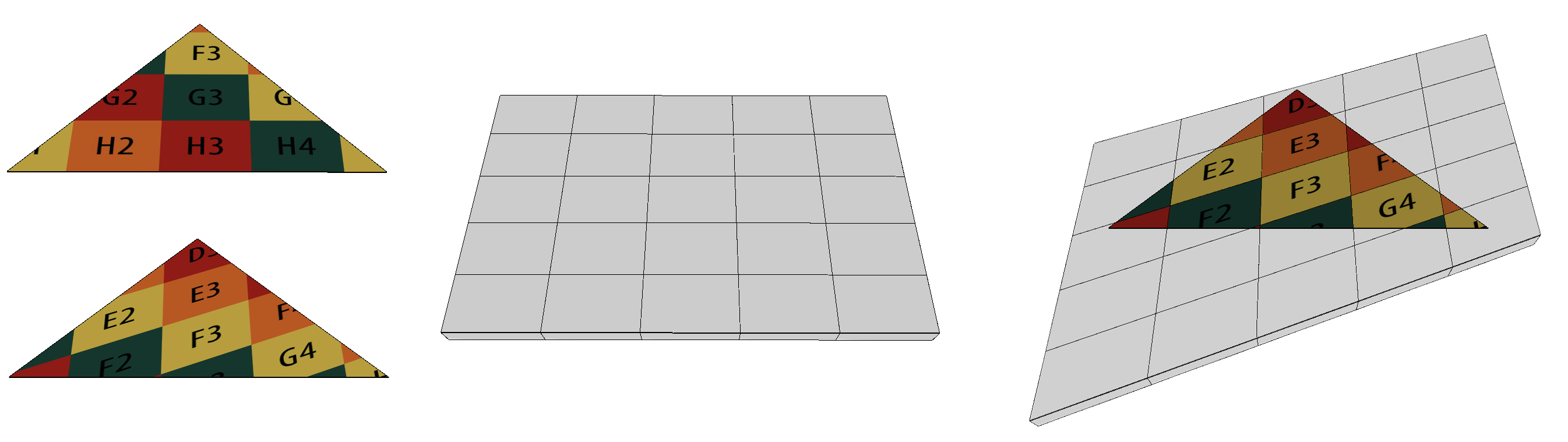}
    \caption{Illustration of a transformation of $m$LAS patch to the texture coordinates space of the triangle. Left-top: textured triangle with aligned $uv$ texture coordinates. Left-bottom: Textured triangle with generic $uv$ texture coordinates. Middle: example of $5\times5$ replication area formed by AABBs of $m$LAS tiles. Right: transformation $M_1$ applied to every box from the replication area. Every AABB is fully aligned with the texturing of the triangle. }
   \label{fig:tile_transformation}
\end{figure}

Finally, we compute the scaling and shearing matrix. When texturing a mesh, various projections can be used. If the mesh is of an arbitrary shape, it is not possible to map a continuous texture without any visible distortion. As an example, wrapping of a sphere in a sheet of paper can be mentioned. 
Therefore, we have to determine the transformation matrix $M_1$ that is applied 
on the projected tile so the tile matches the distorted texturing of the triangle. The distortion ratio is computed based on the difference between the tile's world position $g_{xyz}$ and the world position of its neighbor tiles along $uv$-axis in texture coordinates. The output of this computation are 
scaling and shearing transformation matrices. The resulting $M_1$ transformation matrix is obtained by multiplying translation, rotation, scaling, and shearing matrices. All tiles in the replication area use the same transformation matrix; they only have different translation matrices. The illustration of the process can be seen in \autoref{fig:tile_transformation}.

All tiles have the same size; by using tile AABB and tile transformation matrix $M_1$. We terminate the travesal if there is no ray-AABB intersection. If a hit is found, (\autoref{fig:mambrane_renderer}  (b-c)), we use the tile position in texture $g_{uv}$ to obtain the corresponding Wang tile $Id$ from the tiling recipe that is mapped to the corresponding replication area slot.

As Wang tiles are stored in object space, we test inverse ray intersection with non-transformed Wang tiles. To compute the inverse ray of $a+tb$, we multiply the ray origin and direction with the inverse of the tile's transformation matrix $M_{1}^{-1}$. 

Then, we cast the ray $M_{1}^{-1}a+tM_{1}^{-1}b$ to the $m$LAS, looking for a ray-AABB intersection. If a hit is found at point $p$  (\autoref{fig:mambrane_renderer}  (d)), the traversal continues to the intersected molecule AS at $n$LAS. Otherwise, we continue looking for another intersection in the next tile inside the replication area. For the hit $p$, 
we need to compute the transformation matrix $M_2$ that virtually places the nanostructure in the scene. 

Every molecule $n$ within the Wang tile has a local transformation matrix that represents its position $n_{\mu\gamma}$ and rotation $n_{\theta}$. 
We compute the molecule translation matrix by multiplying the molecule's local position with the tile transformation matrix $M_1$. 

To compute the rotation of the molecule, we use the up vector associated with every type of the molecule. First, we obtain the position of the molecule in $uv$-coordinates by adding the molecule's local position in the tile to the center of tile in $uv$-coordinates. 
Using barycentric interpolation of normals of triangle vertices, we obtain an interpolated normal for a position of the molecule. By multiplying this rotation with the molecule's local rotation $n_{\theta}$, we obtain the resulting molecule rotation. Finally, we then compute the transformation matrix $M_2$ by multiplying the molecule's translation and rotation matrices. If the instance is located outside the prism, that instance intersection is rejected. Otherwise, the traversal continues to the next level in the tree. This early ray termination allows the ray traversal to avoid testing the nanostructures' geometry that falls outside the prism.   

As molecules are stored in $n$LAS in object space, we compute the inverse ray of $a+tb$ by multiplying the ray origin and direction with the inverse of transformation matrix $M_{12}^{-1}$ where $M_{12}$ is the result of multiplying the tile transformation matrix $M_1$ by the molecule transformation matrix $M_2$. Then, we cast the ray $M_{12}^{-1}a+tM_{12}^{-1}b$ into $n$LAS, to detect a ray-sphere intersection as the atoms are the lowest primitive in $n$LAS. If there is a hit, we obtain a hit point $d$ (\autoref{fig:mambrane_renderer} (e)). 

Every atom in the molecule has a local transformation matrix that represents its local coordinate position within the model of the molecule. From this position, we construct the atom transformation matrix $M_3$. As this is the lowest level in the three-level hierarchy, the hit information is recorded and it becomes a candidate for ray traversal output. The hit information is computed for the point $d$ and world transformation matrix $M_{123}$. This matrix is obtained by multiplying $M_1$, $M_2$ and $M_3$.

After the hit candidate is detected, the traversal is not terminated, but the candidate position is used to update the ray $t_{max}$. This value represents the maximum distance of the potential intersections along the ray. By that, the ray traversal ignores any prisms, Wang tiles, and atoms that are further than $t_{max}$. 
\subsubsection{Optimizing the Shell Space Renderer}
\label{sec:opt-replication-area}

In the shell space renderer, several Wang tiles are mapped to each prism (\autoref{fig:mambrane_renderer}  (a-b)). When a ray intersects a prism, it is not known which tile in the replication area has the closest nanostructure to hit. Therefore, we need to traverse all the tiles that intersect the ray and trace it down until their lowest level to identify the closest hit. To optimize this stage, we extend our three-level AS by adding another AS (only for the shell space renderer) between the $\mu$LAS and $m$LAS (referred to as $rep$LAS). The main goal of adding $rep$LAS is to perform ray traversal in the replication area tiles in parallel. We use the replication area predefined dimensions to create a 2D grid of tile's AABBs and place them in object space (see \autoref{fig:tile_transformation}).

The shell rendering algorithm runs with $rep$LAS with a minor adjustment. It starts by shooting the ray $a+tb$ towards $\mu$LAS. Once the ray hits a prism, it computes the transformation matrix $M_1$. It computes rotation, scaling, and shearing matrices exactly the same as explained in the previous section. For the translation matrix, we need to compute the position of the replication area in world space. We employ the texture coordinates of the triangles of the mesh to obtain the center position of the replication area in $uv$ space. Using interpolation, we obtain the corresponding world position and use it for the computation of the transformation matrix $M_1$. We use $M_1$ to inverse ray and shoot it into $rep$LAS. Once a hit is found, we get the Wang tile Id. Then we inverse the ray and shoot it into $m$LAS of that Wang tile. Then, the ray traverses the $m$LAS and $n$LAS as we explained above.

\subsection{Core Space Renderer}
To render the internal parts of the structures, the ray traverses the core grid and identifies grid boxes that intersect with the ray. Either an intersection with the molecular data is found or the ray exits the grid. After a ray-AABB intersection is detected, it idles the traversal on grid space AS and continues the traversal to $m$LAS and $n$LAS. These two levels are processed in parallel. Although it is possible to perform parallel ray traversal on the grid boxes as well, it requires memory allocation for creating an AS for the core grid. We aim not to create nor update AS while rendering. Cells and viruses are typically crowded with soluble components; therefore, in most cases, a hit is found in the first box intersecting with the ray. For this reason we choose to run the grid box traversal sequentially and reduce memory consumption. The extension of the algorithm to run in parallel is straightforward.
\begin{figure}[h]
    \centering
    \includegraphics[width=0.9\linewidth]{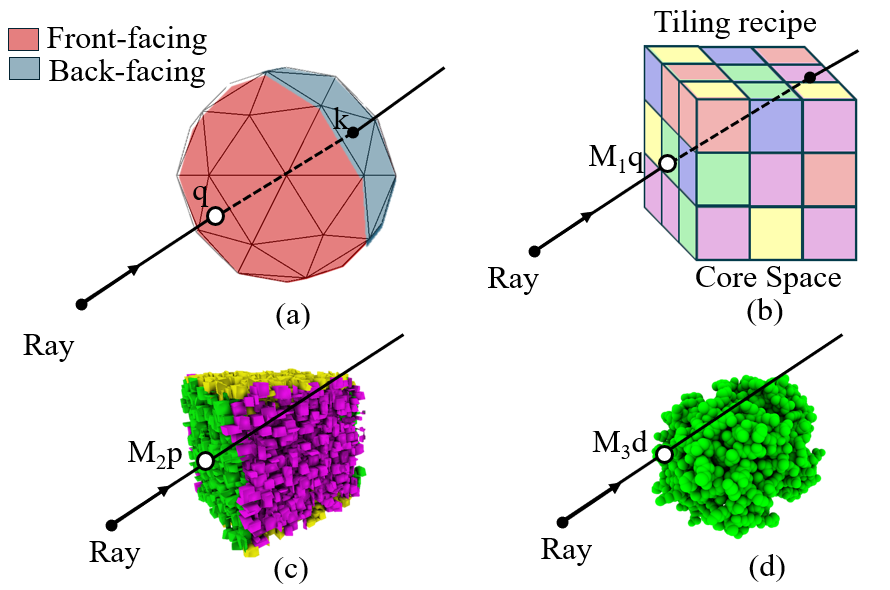}
    \caption{Illustration of Core Space Renderer algorithm.
    }
   \label{fig:soluble_renderer}
\end{figure}
\autoref{fig:soluble_renderer} illustrates the core rendering algorithm. First, we identify a mesh instance that has to be traversed. Second, we continue in its core space and define the ray interval $[t_{min},t_{max}]$. This interval determines the range along the ray that is considered for intersection tests with molecular data. Only intersections within that interval are reported by the ray traversal. The algorithm begins by casting the ray $a+tb$ twice into the $\mu$LAS (refer to \autoref{fig:soluble_renderer} (a)). The first ray identifies the closest front-facing triangle of the proxy geometry intersecting the ray. The second ray identifies the closest back-facing triangle. This results in the hit points $q$ and $k$. If the hit point $q$ is closer to the viewpoint than $k$ (i.e., $t_q < t_k$), the viewpoint is outside of the mesh volume. In this case the ray interval is set to [$t_q, t_k$]. Complementary, if the hit point $k$ is closer to the viewpoint than $q$ (i.e., $t_q > t_k$), the viewpoint is inside the core grid and the ray interval is set to [$\epsilon, t_k$], where $\epsilon$ is a small positive number.

With the start and exit position $q$ and $k$, we compute the coordinates of the start and exit boxes as described in~\autoref{eq:cijk_eq}. Once the traversal interval is defined, we run the ray-grid traversal. Any ray-grid traversal algorithm can be used, for example~\cite{amanatides1987fastvoxeltraversalalgorithm}. Starting from the start box, all boxes intersecting the ray are traversed sequentially until either a hit in the lowest level AS is found or the exit box is fully processed. If the ray-AABB intersection point $q$ is found (refer to \autoref{fig:soluble_renderer} (b)), we obtain the box's coordinates $b_{ijk}$ by exploiting the regularity of the grid. At that point, we stop the traversal at $\mu$LAS and use the 3D tile Wang recipe to look up the corresponding Wang cube to the $b_{ijk}$. We also compute the center position of the box that is used for the computation of the translation matrix for the Wang cube. By multiplying this translation matrix and the transformation matrix of the proxy geometry we obtain the transformation matrix $M_1$. As Wang cubes are stored in object space, we inverse the ray and search for intersection in $m$LAS and consequently in $n$LAS in the same way as described in the shell renderer (refer to \autoref{fig:soluble_renderer} (c-d)).

\subsection{Visibility Management}
Biological models are tightly packed within designated compartments to mimic the crowded environment found in living organisms. This dense arrangement often obscures important internal structures crucial for understanding the organism's functions. Occlusion management techniques address this issue which involves positioning clipping objects within the scene to selectively reveal specific parts of the model. We employ a simple object-space clipping plane that defines the visible region of a scene and determines which elements are rendered based on their spatial relationship to the clipping plane. We apply a trivial reject test at the lowest primitive of each level of three-level AS. At $\mu$LAS, if all vertices of the prism/box fall in the invisible region, that means the mesostructures and nanostructures that should be placed on that prism are also invisible; therefore, in that case, we terminate traversal at that prism/box. On the other hand, if some of the prism's vertices are invisible, that means the prism/box is cut in half with the clipping plane, in this case, we continue traversal at that prism looking for an intersection on its mesostructures and nanostructures levels where the clipping test will be on finer level. 

At $m$LAS, if all vertices of the molecule's AABB fall in the invisible region, we terminate traversal at that level. At $n$LAS, we choose not to apply clipping on the atom level to ensure the accurate representation of molecules in the final scene.

\subsection{Animation}
Living systems are inherently dynamic and complex, which make them challenging to be effectively explained with static representations. Implementing animation on hardware-accelerated raytracing requires frequent updates of transformation matrices or updating the geometry itself. In both cases, the acceleration structure has to be updated or rebuilt. The complexity of this operation rises with the number of elements in the scenes. Updating of AS with millions of elements might take hundreds of milliseconds, posing a significant challenge in handling massive scenes.

Given the nature of our algorithm, integrating specific animation comes at a minimal cost. We integrate two types of animations in our system. 1) proxy geometry transformation and 2) protein instance jittering simulating the continuous optimization process of proteins in living organisms. To integrate proxy geometry transformation, their respective transformation matrices inside $\mu$LAS have to be updated. Typically, we have tens of them and even though update requires rebuilding of $\mu$LAS, this process is fast and does not block the rendering. In this case, Mesostructures and nanostractures automatically adjust to the transformation of the proxy geometry without the necessity of updating their acceleration structures. This is due to the design of our system as the instances are instantiated virtually in our scene. The jittering simulation is implemented by manipulating the instance rotation that is used to compute the transformation matrices $M_2$ based on various parameters, such as time. Consequently, this alteration induces animation effects on the $n$LAS level. This computation is done in the rendering shaders and does not require update nor rebuild of AS. Both types of animations can be seen in the supplementary video.

\section{Results \& Discussion}
Our novel rendering approach is capable of on-the-fly constructing and visualizing massive biological worlds with minimal memory footprints.
We demonstrate the capability of Nanouniverse on two scenes containing tens of RBCs and SARS-COV-2 virions.

We use a set of 16 tiles configurations of Wang square tiles for each model. For solubles, we use 16 Wang cube tiles for the RBC model and 8 predefined cubes with complex non-repetitive geometry for SARS-CoV-2. The animated scene fly-through with clipping plane support can be seen in the supplementary movie.

To provide insight about the performance, we tested the system on AMD Ryzen Threadripper PRO 3975WX processor with RTX 4090 GPU, with one ray per pixel when measuring the framerates. The system is implemented using C++/17 with Vulkan API and NVIDIA's nvpro-samples framework~\cite{nvpro-samples}. We use {\it GLSL\_EXT\_ray\_query} extension with compute shader in order to launch a new ray query during ray traversal. 
For the evaluation, we used two models - SARS-CoV-2 and RBC. Fully populated SARS-CoV-2 consists of approx. 24M of atoms, whereas fully populated RBC consists of approx. 1.2 trillions of atoms. We use both models in the two test scenes. The first scene consists of 4 RBCs and 20 SARS-CoV-2 virions, and the second scene contains 20 RBCs and 40 SARS-CoV-2 elements. 
In both scenes, these biological entities are randomly placed and oriented. The performance of the algorithm together with the memory consumption in both scenes can be found in \autoref{tab:performance}. It is important to highlight that the algorithm is memory lightweight. Because the whole scene is constructed on-the-fly from basic building blocks, only limited amount of memory is required. More specifically, we upload approx. 103MB of data (low-poly proxy geometry, positions and rotations of proteins within Wang tiles and atoms of protein structures). Raytracing itself allocates approx. 373MB of supporting acceleration structures for our data. 
We reached interactive rates allowing the user to freely explore the provided scenes yet still preserving fully atomistic details. 

\begin{table}[t]
\centering
\caption{Performance of the algorithm}
\begin{tabular}{p{0.15\linewidth}p{0.12\linewidth}p{0.10\linewidth}p{0.08\linewidth}p{0.08\linewidth}p{0.08\linewidth}p{0.08\linewidth}}
\hline
 Resolution & \# SARS & \# RBC & min [FPS] & avg [FPS] & Data [MB] & AS [MB]\\
\hline
1920x1080 & 20 & 4 & 19 & 24 & 103 & 373 \\
3840x2160 & 20 & 4 & 9 & 11 & 103 & 373 \\
1920x1080 & 40 & 20 & 11 & 14 & 104 & 373 \\
3840x2160 & 40 & 20 & 7 & 8 & 104 & 373 \\
\hline
\end{tabular}
\label{tab:performance}
\end{table}

\begin{figure}[h]
    \centering
    \includegraphics[width=\linewidth]{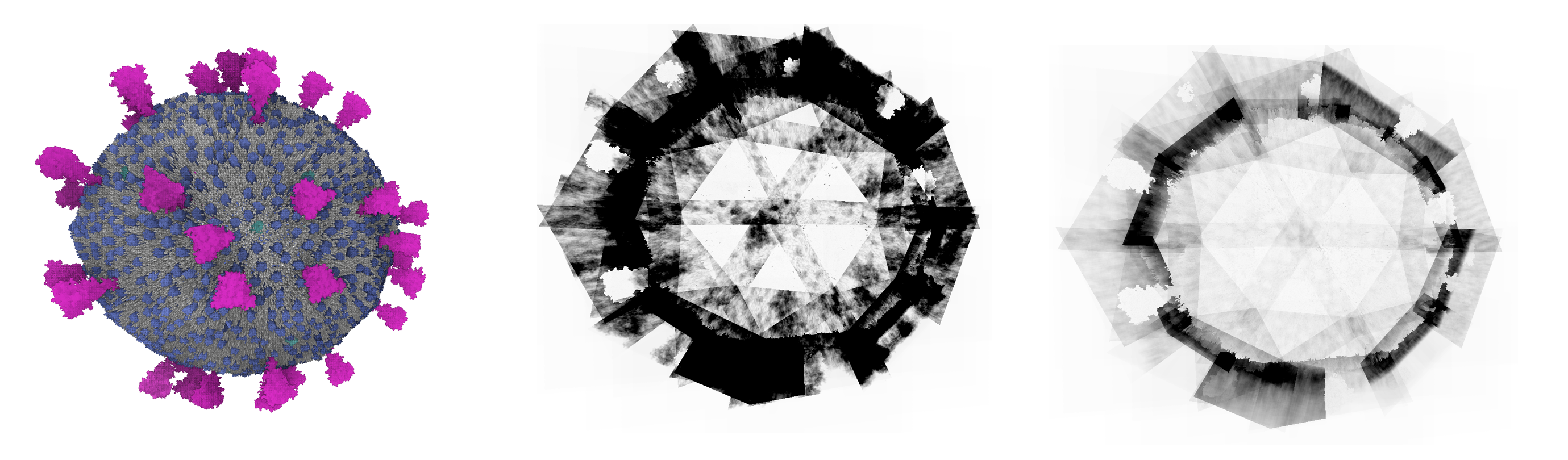}
    \includegraphics[width=\linewidth]{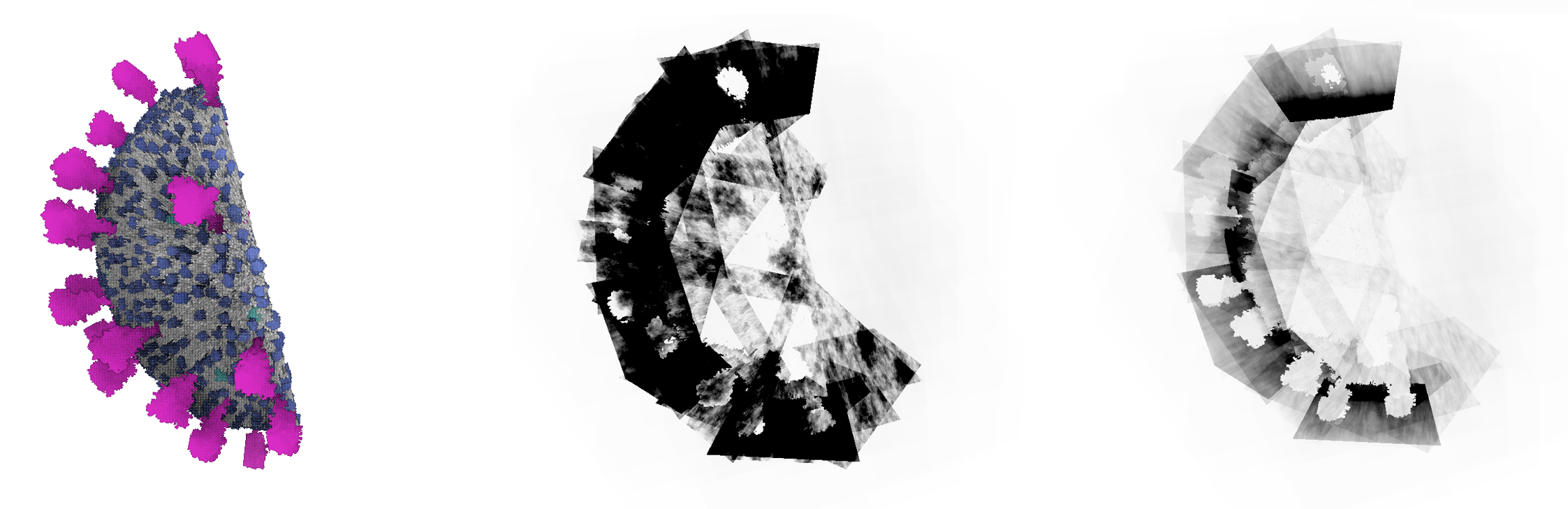}    
    \caption{Side-by-side comparison showing the number of intersection tests that are performed to render the membrane of SARS-COV-2 (left) using two different AS designs. The heatmap ranges between 0 hit (white) to 300 hits (black). Middle: two-level AS with prisms forming the top level and both mesostructures and nanostructures in the bottom level. The performance is 33 (top) and 41 (bottom) FPS. Right: the proposed three-level AS with the performance of 126 (top) and 172 (bottom) FPS.}
   \label{fig:3vs2levelAS}
\end{figure}

\begin{figure}[h]
    \centering
    \includegraphics[width=\linewidth]{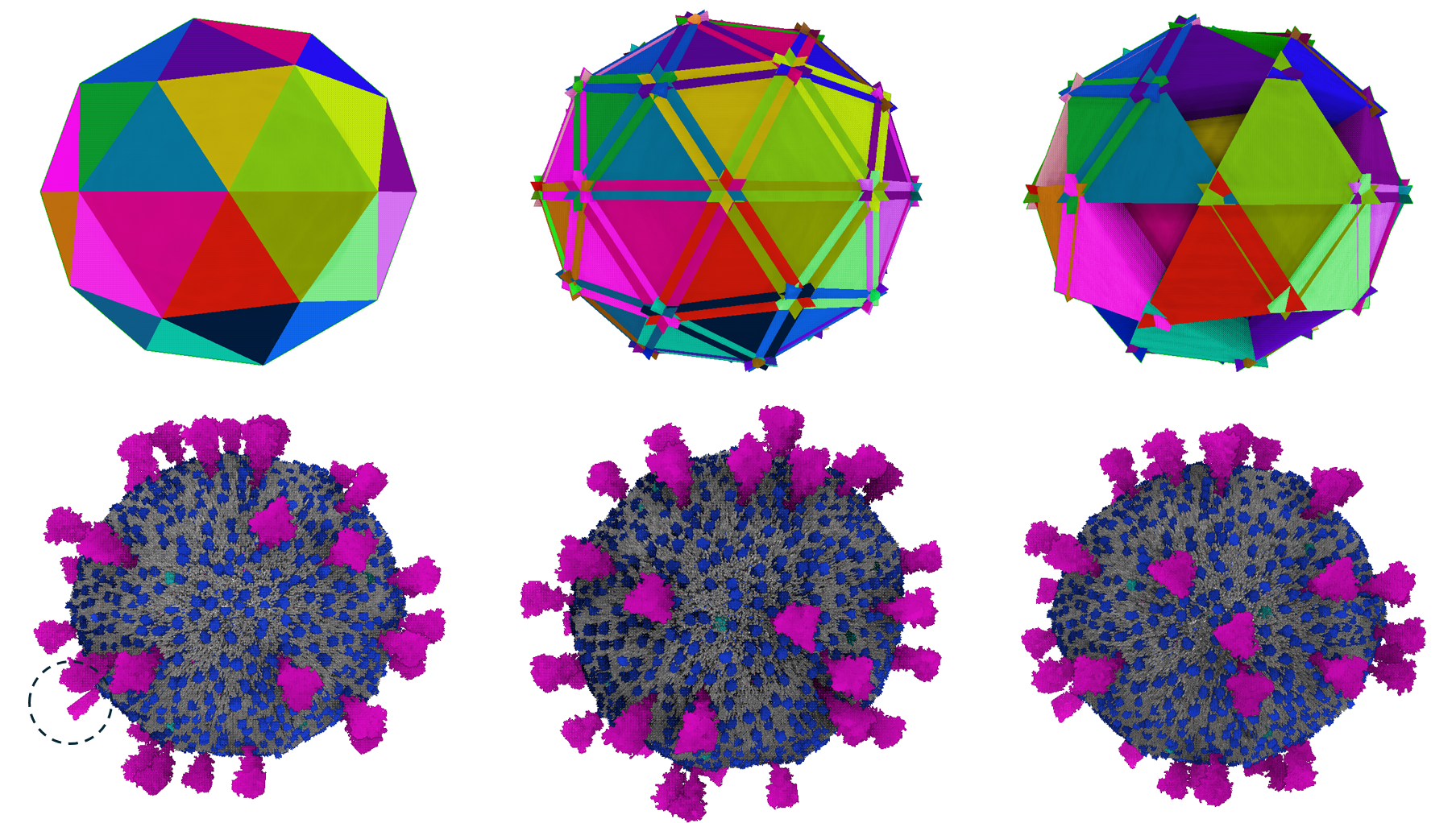}
    \caption{Rendering result with the different shell space design (half missing protein highlighted with dotted circles) . From left to right, prisms with equal height, prisms with equal height and overlaps,  prisms with adaptive height.  }
   \label{fig:adaptiveshell-result}   
\end{figure}

\begin{figure}[h]
    \centering
    \includegraphics[width=\linewidth]{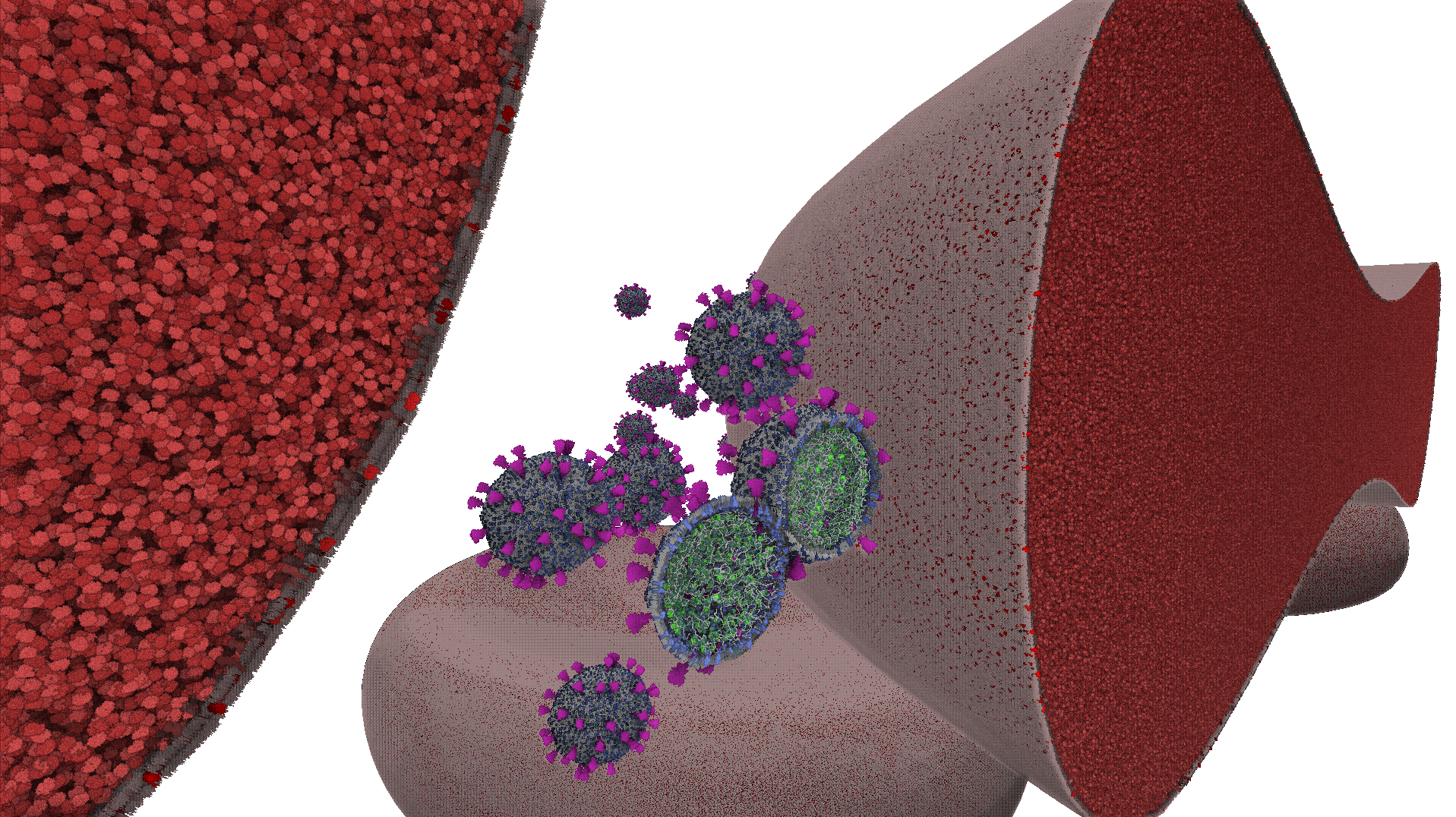}
    \caption{Rendering of several SARS-CoV-2 from a cut open RBC.}
   \label{fig:result-cut-open-RBC}   
\end{figure}

We have tested several designs for acceleration, and we reported the one that gives the best result. For example, we tested two-level AS with prisms forming the top level and both mesostructures and nanostructures forming the bottom level. \autoref{fig:3vs2levelAS} shows a side-by-side comparison between the two-level AS tree and three-level AS tree on rendering the membrane of SARS-COV-2. We use a heatmap to show the number of interested tests performed by each design. The heatmap ranges between 0 hit
(white) to 300 hits (black). The heatmap shows how the proposed three-level AS tree reduces the number of intersection tests.

\autoref{fig:adaptiveshell-result} shows the results of different shell space designs. The leftmost image displays prisms with equal heights, while the corresponding image at the bottom demonstrates the visual result rendered at 73 FPS. Some spike proteins have half of their parts missing, because of their placement at the edge of the instance. Increasing the width of the prisms, causing them to overlap, ensuring the correct rendering of the entire mesostructures and nanostructures placed on them. The overlapped prisms with equal heights are depicted in the middle, with the corresponding image at the bottom showing the visual result rendered at 56 FPS. Clearly, there is room for improving the performance by skipping the empty space in the shell layer. A straightforward approach is to assign a different offset to each triangle as needed, allowing the shell space to adapt its size based on the contained mesostructure. The rightmost image illustrates the adaptive prisms with overlap, while the corresponding image at the bottom depicts the visual result rendered at 70 FPS. Utilizing the adaptive shell without overlaps increases performance to 90 FPS. It would be interesting to integrate empty space-skipping techniques from volume rendering literature to accelerate the ray traversal within the shell space.

\section{Conclusion}
Given the substantial difference in size between cells and their components, spanning over seven orders of magnitude, numerous efforts have been made to achieve seamless visualization across diverse structural levels within the biological mesoscale. This paper represents a step toward addressing this challenge.

The algorithm is general enough that it can be used for different types or modalities of data, for example generating cities, forests or similar heavily instance-based populated scenes. Moreover, the whole concept can be extended to higher amount of levels than three. 

The system supports levels-of-detail even though we have not used it in this paper as we aimed to visualize full details. This is a straightforward extension as the next traversed AS is determined during the run-time. It is simple to point towards the lower AS with low details during this phase.

In the future, we would like to explore the extensibility of the system. Our next steps will lead towards more complex animations. For example, we can have multiple timesteps of a single Wang tile as a set of keyframes. Later, in the rendering phase, these keyframes can be used for direct interpolation. We believe that this approach might provide a platform to visualize (and animate) more complex processes such as proteins production or proteins interaction. 

Another extension can be procedural generation of the scene. As of now, we specify amount, positions, and rotation of proxy geometries manually. If this is generated automatically, the system can offer an unlimited exploration space.

Further focus in the future will be on optimization of the system. We expect that integrating the latest RTX features together with other NVIDIA technologies such as DLSS will allow us to bring the whole system to AR/VR for educational and presentation purposes.

\section*{Acknowledgment}
The research was supported by the King Abdullah University of Science and Technology (BAS/1/1680-01-01).

\bibliographystyle{unsrt}
\bibliography{main}

\end{document}